# DotGrid: a .NET-based cross-platform software for desktop grids

## Alireza Poshtkohi*

Department of Electrical Engineering
Qazvin Azad University
Daneshgah St., P.O. Box 34185–1416, Qazvin, Iran
E-mail: alireza.poshtkohi@gmail.com
*Corresponding author

## Ali Haj Abutalebi and Shaahin Hessabi

Department of Computer Engineering
Sharif University of Technology
Azadi St., P.O. Box 11365–8639, Tehran, Iran
E-mail: abutalebi@ce.sharif.edu
E-mail: hessabi@sharif.edu

**Abstract:** Grid infrastructures that have provided wide integrated use of resources are becoming the *de facto* computing platform for solving large-scale problems in science, engineering and commerce. In this evolution, desktop grid technologies allow the grid communities to exploit the idle cycles of pervasive desktop PC systems to increase the available computing power. In this paper we present DotGrid, a cross-platform grid software. DotGrid is the first comprehensive desktop grid software utilising Microsoft's .NET Framework in Windows-based environments and MONO .NET in Unix-class operating systems to operate. Using DotGrid services and APIs, grid desktop middleware and applications can be implemented conveniently. We evaluated our DotGrid's performance by implementing a set of grid-based applications.

**Biographical notes:** Alireza Poshtkohi received his BSc degree in Electrical Engineering from the Islamic Azad University of Qazvin, Qazvin, Iran in 2006. He is currently a Grid Computing Researcher at the Islamic Azad University of Qazvin. His current research interests include grid computing and distributed computing in .NET platform.

Ali Haj Abutalebi received his BSc and MSc degrees in Electrical Engineering from the University of Tehran, Tehran, Iran in 1995 and 1998, respectively. He is currently a Lecturer at the Sharif University of Technology. His current research interests include system-level design and verification, reconfigurable systems and distributed computing.

Shaahin Hessabi was born in Tehran, Iran on 14 February 1961. He received his BSc and MSc degrees in Electrical Engineering from the Sharif University of Technology, Tehran, Iran in 1986 and 1990, respectively. He received his PhD degree in Electrical and Computer Engineering from the University of Waterloo, Waterloo, Ontario, Canada in 1995. He joined the Sharif University of Technology in 1996, and is currently an Associate Professor at the Department of Computer Engineering. His current research interests include testing and design for testability, verification, VLSI design, SoC and NoC, and reconfigurable systems.

## 1 Introduction

Grid infrastructures (Foster *et al.*, 2001; Foster *et al.*, 2002) provide the ability to share, select and aggregate distributed resources as computers, storage systems or other devices in an integrated way. Grid has solved many problems in science, engineering and commerce fields. This has come true via implementing various middleware technologies (Foster, 2006; Thain *et al.*, 2003; Grimshaw and Wulf, 1997; Huber, 2001; Buyya and Venugopal, 2004; Luther *et al.*, 2005) for grid. In the grid area, the desktop grid technologies are sought for an interesting field to harness the computing power of desktop computers. Desktop grids are attractive platforms for running compute-bound distributed applications as they can provide low-cost, otherwise unused CPU cycles and make it possible for enterprises to maximise return-on-investment for desktop resources. Desktop grid technologies allow the grid communities to exploit the idle cycles of pervasive desktop PC systems to increase the available computing power. Many researchers have investigated this area and developed desktop grid software. Equally, regarding the fact that most power of desktop PCs in organisations and home environments are underutilised, many enterprise desktop grid softwares have been created by commercial enterprises (Sullivan *et al.*, 1997; Anderson *et al.*, 2002; Chien *et al.*, 2003; United Devices, 2007; Platform Computing, 2007). The key issues that must be considered in a grid infrastructure are heterogeneity, reliability, application composition, scheduling, resource management, security and high-throughput data transfers. In this paper we introduce a cross-platform grid infrastructure to implement desktop grid systems, called DotGrid (Poshtkuhi *et al.*, 2006; Abutalebi *et al.*, 2006). DotGrid platform is the first comprehensive desktop grid software utilising Microsoft's .NET Framework (Microsoft Corporation, 2007) in Windows-based environments and MONO .NET (2007) in Unix-class operating systems to operate, and provides a set of standard interfaces to co-exist and interoperate with widely used grid middlewares such as Globus, Condor, Legion and UNICORE. DotGrid tries to set up a cross-platform grid toolkit including grid services and APIs required for implementing grid middleware and applications aiming at desktop grids. DotGrid is implemented as a layer over the existing OS. This layered approach eliminates the dependency of grid to the native OS. Hence, all middleware and applications can be developed in such a grid regardless of the type of the hardware and software over it, because the DotGrid layer provides all needed services and system information needed for middleware and grid applications. Developing this layer leads us to implementing a homogenised system that is required for a grid infrastructure. DotGrid proposes an implementation for grid based on Microsoft .NET platform. Although .NET is introduced in Windows OS, it has now become a multiplatform environment through some projects such as MONO .NET that have made this platform available in other operating systems, *e.g.*, Linux, Solaris and other UNIX-based OSs. The other reason that biases .NET platform as our choice is that its rich library bypasses the need for implementing numerous required grid services from scratch. Microsoft has also standardised .NET platform and its languages at ECMA (ECMA-334, 2007; ECMA-335, 2007). MONO .NET in Linux and Microsoft .NET in Windows together provide a cross-platform .NET where a wide range of machines regardless of their OS may build up a grid.

In comparison to other grid middleware, the cross-platform feature of DotGrid leads to true universal grid infrastructure and bypasses incompatibility problems. DotGrid is implemented as a layer above the .NET platform. Figure 1 shows DotGrid structure. Figure 2 shows current developed DotGrid services and APIs. The rest of the paper is organised as follows. Section 2 describes related works. Section 3 briefly describes DotGrid architecture and Section 4 presents some applications developed based on DotGrid. In Section 4.2, Alchemi middleware and DotGrid are compared to each other by solving a typical computational application developed based on Alchemi and DotGrid.

**Figure 1** DotGrid structure

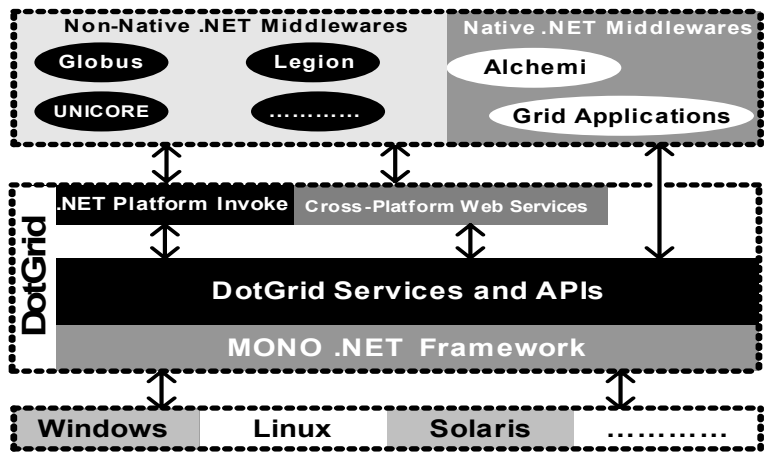

**Figure 2** Current developed DotGrid services and APIs

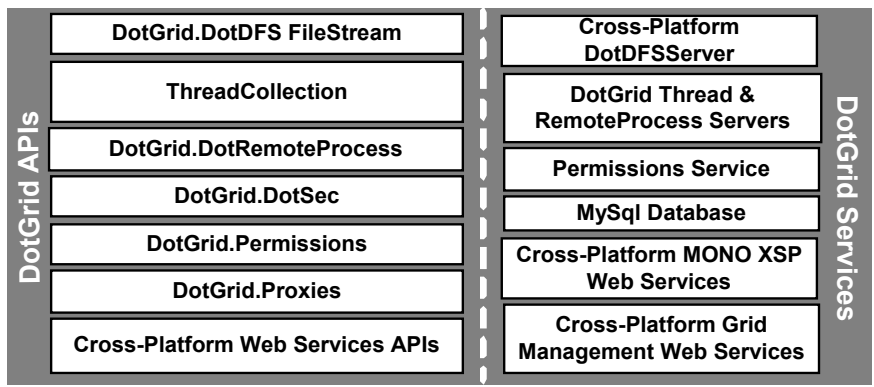

## 2   Related works

Recently, some .NET-based grid computing tools and middleware have been created. Alchemi middleware at the University of Melbourne is an open source software framework based on .NET that allows to painlessly aggregate the computing power of networked Windows machines into a computational grid (Luther *et al.*, 2005; Alchemi, 2007). WSRF.NET, created at the University of Virginia, is a .NET-based hosting environment for grid services that implements Web Services Resource Framework (WSRF) and supports the Open Grid Services Architecture (OGSA) on .NET Framework (GCG, 2007).

## 3   Architecture

DotGrid benefits from a layered and service-oriented architecture and provides the necessary services and toolkits that are required for building grid, cluster and peer-to-peer computing environments. DotGrid services, which are executed as servers (daemons), run in every operating system such as Windows, Linux and Solaris and provide a dynamic runtime environment. This way, grid or cluster middleware can be developed easily in heterogeneous or homogeneous distributed environments. Based on the DotGrid toolkit, one can interface his/her systems to other existing grid middleware such as Globus via .NET Platform Invoke and cross-platform web services.

### *3.1   DotGrid.DotDFS service*

DotDFS is a grid-based High Throughput File Transfer System that is used as an infrastructural framework enabling data grid characteristics (Belalem and Slimani, 2007) for resource sharing and files transferring in DotGrid platform. DotDFS proposes a set of open and cross-platform binary protocols. Most DotGrid services use DotDFS. For brevity, we only focus on the DotDFS protocol, and not the core service implementation.

#### *3.1.1   DotDFS protocol*

DotDFS is a flexible protocol that provides a substantial infrastructure for resource sharing and high-throughput file transfer on current grid substructures and on our DotGrid platform. DotDFS protocol introduces two operating modes, and clients need to choose one of them while making a connection to the DotDFS server. One of these modes must be selected based on demand of the developed grid application:

- Distributed File System Mode (DFSM)

  In this mode of operation, DotDFS server provides distributed files accessing and data sharing mechanism services similar to those provided by conventional distributed file systems. These services are based on the Grid instinct requirements such as security and web services. The DotDFS protocol in this mode of operation that actually exhibits a grid-based distributed file system behaviour can be used as data file stripping mechanism (Poshtkuhi *et al.*, 2006) that exists in GridFTP protocol (Allcock, 2003; Allcock *et al.*, 2005a–b). DotDFS supports *Read*, *Write*, *Flush*, *Lock*, *UnLock*, *Seek*, *Close* and *SetLength* file system operations. Applications that require different file systems can be distributed conveniently on grid computing environments using this infrastructure. This protocol supports both random and sequential access to distributed files.

- File Transfer System Mode (FTSM)

  This mode of operation, like the GridFTP protocol, includes mechanisms through which clients can transfer data between a single source and destination using parallel TCP connections for improving and aggregating the bandwidth in comparison with single TCP stream.

More features of DotDFS protocol are as follows:

- Reliable transfer of data files

  Reliable transfer of data files is one of the vital objectives of grid applications in grid computing. The DotDFS protocol defines procedures that using them and available API implementations at client-side make reliable and confidential data files transfers on the network possible. The DotDFS protocol retransfers network packets that had been transferred in unreliable network situations to the opposite side by predefined policies.

- Parallel file access

  The DotDFS protocol allows parallel access to a single file system from unlimited number of DotDFS clients in DFSM mode. The .NET file lock can prevent collisions and ensures data consistency. Parallel file access accelerates processing of grid applications that share data; this way multiple copies of data is not needed and data storage space is reduced. This approach is suitable for some data-intensive grid environments such as high-energy physics applications.

- Partial file transfers

  Partial file transfer means that it is possible that grid processes and applications tend to transfer only fragments of a file. This mechanism is supported by defined DotDFS headers.

- Stateless architecture

  The DotDFS protocol introduces a natural stateless architecture in DFSM operating mode. This states that DotDFS servers do not keep track of DotDFS client requests according to which files have been opened, file positions, *etc*. This approach leads DotDFS clients be able to fail and resume without disturbing our system as a whole and as a result this mechanism allows fast developments of grid-based parallel cluster file systems, in the future.

- Transfer security modes and network stream buffer sizes

  The DotDFS protocol specifies three modes for transmitting data files on networks: secure, non-secure and semisecure modes. In secure mode, all data are encrypted and then transferred on network, but in non-secure mode, only user credentials and file handles or file path names are encrypted according to the DotSec protocol (Poshtkuhi *et al.*, 2006; Abutalebi *et al.*, 2006). For more flexibility, DotDFS protocol introduces semisecure mode in which grid developers can use secure and non-secure modes simultaneously in order to improve performance. The DotDFS protocol provides some headers for TCP Buffer/Window sizes that are transferred between clients and servers in the beginning of any DotDFS session. When the best TCP Buffer/Window size is selected, high-throughput data transfer in parallel TCP/IP transfers can be accessed specially on WAN networks.

- Authentication and data integrity

  Authentication and data integrity are two important objectives for securing DotDFS file transfers. DotDFS uses DotSec mechanisms for secure authentication, user authorisation and preserving data file integrities.

### 3.2 *DotGrid.DotSec security layer*

DotSec (Poshtkuhi *et al.*, 2006; Abutalebi *et al.*, 2006) that is designed based on the DotDFS architecture for increasing throughput and bandwidth usage during file transmission in secure mode, is used in the DotGrid platform for secure communications and user credential authentications.

### 3.3 *DotGrid.DotThreading service*

DotThreading service causes applications that want to use threading for parallelism and want to distribute threads in a network, execute their threads remotely in a grid or cluster environment and get the results. DotThreading service consists of one class API named ThreadCollection at the client side and a server named DotGridThreadServer. Figure 3 depicts mechanisms based on which a client runs a set of threads over a remote DotGridThreadServer session. In this figure, just the DotGrid services and APIs that are involved in this scenario are shown and the remaining and services and APIs are dropped.

- ThreadCollection client API

  This class API is responsible for creating and managing n remote threads on a remote thread server. Authentication is performed once for all of these n threads on the server. Creating a set of threads via this class is equivalent to a process that has been remotely executed. This client API has secure and non-secure modes for data transfer between clients and servers based on the DotSec Model.

- DotGridThreadServer server

  This server is responsible for receiving and executing client API requested threads. This server, like the DotDFS server, is assigned to client authentication and management of client consumed system resources. To understand this server architecture, let us study the scenario shown in Figure 3. In this figure, ❶ a client located at 'node i' contacts DotGridThreadServer located at 'node j' by using ThreadCollection class, remotely. For contacting, client mentions items such as method delegates (similar to function pointers in C language) for a remote run 'execution', module dependencies and user credentials. After authenticating, serialised data of classes that are to be executed remotely are transferred to 'node j'. ❷❸❹ At this time 'node j' peer is connected to 'node i' peer via DotDFS FileStream API and DotDFS downloads the files that have been declared by the client in 'node i' as module dependencies to local hard disk of 'node j'. ❺ Then

'node j' creates a .NET Application Domain and after configuring its related permissions, starts executing the client threads and finally results are returned back to client in an array of objects.

**Figure 3** DotThreading architecture and interaction between DotGrid services and its APIs in a peer-to-peer scenario

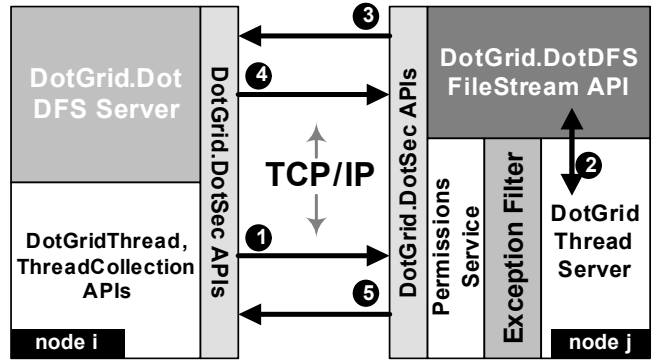

## 3.4 DotGrid permissions service

Management of complex, distributed, and dynamically changing job executions and resource usages are some central problems in grid environments (Keahey *et al.*, 2003). All modern operating systems declare an elaborated set of permissions in order to handle users' level of access to system resources and enforcing local policies. Regarding that DotGrid is an infrastructure that intends to develop cross-platform grid middlewares, it includes a service named File Permissions Service (FPS) (Abutalebi *et al.*, 2006) designed for managing users' distributed file system access level and enforcing grid policies. DotGrid uses FPS for enforcing all processes to use DotDFS File Stream APIs and Virtual Directory Services (VDS). DotGrid uses .NET Code Access Security (CAS) model for investigating system resources access permissions during remote code runtime execution of services such as DotThreading and DotRemoteProcess. Based on the CAS model, each application that uses .NET Common Language Runtime (CLR) for execution must interact with runtime's system security. At the time of an application execution, system security investigates the permission set declared by CAS. If that application is using a resource that is not permitted by CAS, the runtime's system security throws a security exception based on predefined DotGrid permission policies at that application domain's space. This way, that application will not be able to use that forbidden resource(s). These permissions have been set in application domain's space of the related grid application before run. Enforcement of these CAS permissions has only been applied to native .NET applications that use DotThreading or DotRemoteProcess services. For non-native .NET job submissions based on DotRemoteProcess that execute grid jobs on remote nodes on Linux or Windows OS, these mechanisms cannot be applied; in these cases DotGrid can only prevent the native code execution before run. DotGrid defines two types of user accounts for enforcing permissions: 'Administrator' and 'Others'. There is no limit for 'Administrator' account application execution for using system resources. For 'Others' account, imperative permissions is applied. In Figure 4 typical imperative permissions for an 'Others' account is depicted. This information is saved in MySql database and is formatted in XML format.

**Figure 4**   Sample XML-based DotGrid permissions representation

```xml
<?xml version ="1.0" encoding="utf-8"?>
<permissions AccountType="Others">
    <UnmanagedCode value ="True">
Ability  to call unmanged code.
</UnmanagedCode>
    <SocketPermission value ="False"/>
    <Execution value ="False"/>
    <FileIOPermission value ="False">
Controls the ability to access files and
folders.</FileIOPermission>
     <RegistryPermission value="False"/>
     <SqlClientPermission value="True"/>
</permissions>
```

### 3.5   .NET platform invoke service

One of the most effective powers of the Microsoft .NET platform is its capability of interoperating with unmanaged code via Platform Invoke service relying on metadata to locate native exported functions and marshal their arguments at run time (Clark, 2003; Interoperating, 2004). The .NET managed code calls dynamic link libraries such as DLLs (.dll files) in Windows and shared objects (.so files) in Linux through P-Invoke. For example, some managed Microsoft .NET and MONO .NET class library APIs are implemented by direct calling native Windows (Win32 or Win64) and Linux kernel APIs through this mechanism. The C# language function declaration shown in Figure 5 would invoke the POSIX *getpid()* system call on platforms that have the 'libc.so' library. Hence grid middlewares and applications developed based on DotGrid may use APIs and services of other native middlewares such as Globus, developed based on standard C and C++ compilers.

**Figure 5**   Interoperating with a native POSIX function from DotGrid platform through .NET platform invoke and C# language

```csharp
using System;
// for DllImport
using System.Runtime.InteropServices;

[DllImport ("libc.so")]
private static extern int getpid ();
```

### 3.6   *DotGrid.DotRemoteProcess service*

A DotRemoteProcess contains an embedded server in DotGridThreadServer and related client APIs. The embedded server in DotGridThreadServer increases performance and decreases memory usage of any DotGrid instance. The DotRemoteProcess extends .NET Process class and adds remote process creation and management capabilities. One of DotRemoteProcess's features provides mechanisms for remote job submission based on the nature of the process for running on a special operating system (native .NET jobs or non-native jobs like native processes written in C language). In this case DotRemoteProcess creates and manages a remote process and DotDFS transfers input, output and application programme files between two nodes.

### 3.7 Cross-platform web services

DotGrid uses MONO ASP.NET Runtime for cross-platform web services (Baghdadi and Al-Rawahi, 2007) in two ways: one, the mod_mono Apache module that is used to run ASP.NET applications within the Apache web server; and another, MONO XSP which is a HTTP web server written in C# language that can be used to run stand-alone ASP.NET applications. MONO XSP has been recompiled internally in our system kernel. Some parts of DotGrid APIs such as DotDFS FileStream and Cross-Platform Grid Management Web Services are implemented based on web services allowing other grid platforms to interact with DotGrid. Using MONO XSP, projects such as OGSI.NET (GCG, 2007) can also become a cross-platform.

### 3.8 DotGrid summary

Such an infrastructure stated in the above sections causes forming of the architecture of networks with different deployment topologies such as master-slaves, hierarchical and complete graph (currently, the complete graph topology is realised with DotDFS service that provides resource sharing in DotGrid platform among all grid nodes). These topologies are depicted in Figure 6. Such topologies are also discussed in PVM (El-Rewini and Abd-El-Barr, 2005) application developments. Furthermore, DotGrid intrinsically supports development of applications that are based on Single Programme, Multiple Data (SPMD) and Multiple Programme, Multiple Data (MPMD) architectures with its remote grid threading and process model that are DotThreading and DotRemoteProcess services.

**Figure 6** Some typical DotGrid deployment topologies

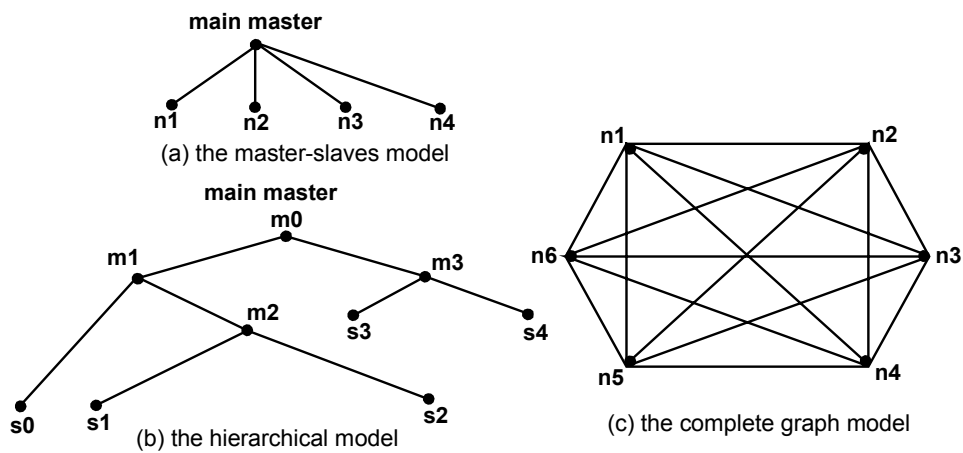

## 4 Performance evaluations

### 4.1 DotGrid.CryptEngine

It is conspicuously important to reach a high performance and high throughput cryptographic engine in internet and distributed environments such as grid and cluster where computers interact with each other. Cryptography is used widely in internet and different networks with transport security layers such as TLS (Dierks and Allen, 1999), GSI (Linn, 1997) and DotSec. Cryptographic systems are divided into three main groups: Asymmetric, Symmetric and Quantum Cryptography Systems. Asymmetric and Symmetric systems are used more widely. All encryption systems use 'keys' for plaintext encryption and converting them into cipher text. The longer the keys are, the longer time of processors will be dedicated to data encryption and decryption. Efforts for improving encryption performance have been done so far and some solutions are proposed. For instance, (Dongara and Vijaykumar, 2003) from ICBC uses multiprocessors that are dependent on hardware and algorithm enhancement. The other effort is done by (Burke *et al.*, 2000). CryptEngine is well implemented in grid environments. Alchemi introduces GridCrypt (Setiawan *et al.*, 2004) in this regard. But in grid environments petabytes of data are generated, thus an engine based on the nature of grid environments must be designed and implemented so that encrypting and storing petabytes of generated data with high performance and throughput would still be possible while grid environments and their data are scaled into peta levels. In such a system, files of high importance can be encrypted, stored and transferred with high throughput equal to ordinary files. In this section DotGrid.CryptEngine that is a grid-based CryptEngine is implemented as one of the subapplications of DotGrid platform, using its services and APIs such as DotDFS, DotSec and DotThreading. We had succeeded before to reach up to 623 Mbps speed in 1000 MB file transmission in a LAN network with 1 Gbps bottleneck, while this throughput in secure mode utilising DotGrid.DotSec for encrypting data files between client and server descends to 106 Mbps (Abutalebi *et al.*, 2006). By implementing CryptEngine, data files can be stored encrypted and there will be no need for DotSec or anything like that for transmitting files between storage systems. At the same time, using this system, sliced large data files encryption in a Cluster File System becomes possible with striping mechanisms. This system can be used for Symmetric and Asymmetric encryption/decryption of very large data in scale of petabytes in a grid or cluster environment with high throughput and performance. Right now, CryptEngine supports Rijndael (Daemen and Rijmen, 2002), 3DES (CryptographyWorld.com, 2006), RC2 (Rivest, 1998) and RSA (Jonsson and Kaliski, 2003) algorithms available in Microsoft .NET Framework. Furthermore CryptEngine provides an abstract class where different cryptographic algorithms such as Blowfish, IDEA, Mars, RC4/6 and Twofish can be added.

#### 4.1.1 Architecture

The main idea in CryptEngine design is to slice an n byte file that is BlockSize long and to distribute cryptographic tasks in a grid environment with m nodes based on DotGrid platform. This slicing mechanism is depicted in Figure 7. Considering experimental results, we chose BlockSize to 20 Mbytes. 'PartNum' is a 64-bits long digit that represents the i-th block of corresponding file. This number is set to be 64 bits length for handling petabytes long files. 'length' represents how long encrypted data is. CryptEngine uses MD5 Hash for keeping the integrity of pure data. CryptEngine has been implemented and tested based on hierarchical master-slaves model of DotGrid deployment topology described in Section 3.7. Figure 8 depicts this architecture in a cluster of Windows XP nodes. The distributor in Figure 8 is responsible for distributing and managing cryptographic tasks among other nodes. The distributor calculates the number of threads that needed remote threads based on file size, *i.e.*, FileSize divided by BlockSize (FileSize/BlockSize) that represents the size of cryptographic tasks. During this time, CryptEngine will instantiate an array with the size of calculated number of threads of CryptInfo class. Each instance of CryptInfo class contains information such as cryptography parameters, DotDFS shared on distributor filename, the IP address of nodes where this thread will store encrypted blocks via DotDFS FileStream API and the length and offset of the i'th block that will be encrypted with this thread. After performing these initial phases, the Distributor node located at node 1 contact with all nodes and sends to DotGirdThreadServer of all nodes the corresponding binary serialised information and then remotely starts all the threads. Figure 9 shows the interaction between the Distributor node and the i'th node while executing encryption. In this figure just the APIs and services of DotGrid that are used in CryptEngine are mentioned. Each node starts encryption/decryption process after receiving its own information. Each thread is connected to DotDFS server of Distributor node via DotDFS FileStream API and in a loop with iteration size equal to size of its DataSize/BlockSize for cryptographic operations, reads streams of files from offset to length (specified in CryptInfo class information) using 'remoteReader' file pointer mentioned in Figure 9. After performing encryption/decryption any blocks, the encrypted/decrypted data block is written on the corresponding node's storage system regarding whether they have to be transmitted to distributor or to be saved in other nodes (*e.g.*, Collector node). This mechanism can be used for saving encrypted mode files in file cluster systems. Through all these steps, each remote thread uses Read, Write, Lock and Seek methods of the related DotDFS for reading, writing and moving the 'remoteWriter' or 'remoteReader' file pointer to the mentioned offset for reading or writing the i'th block to the length of 'length' for encryption/decryption operations.

**Figure 7** CryptEngine header structure for i'th block of a file

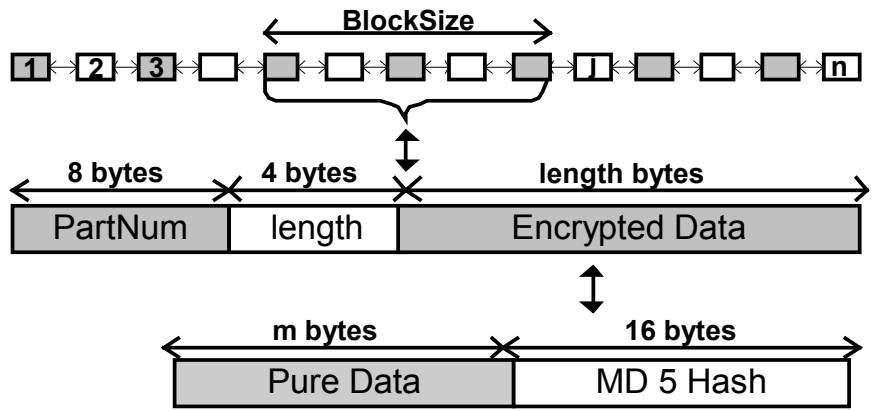

**Figure 8**   The CryptEngine hierarchical master-slaves deployment topology testbed

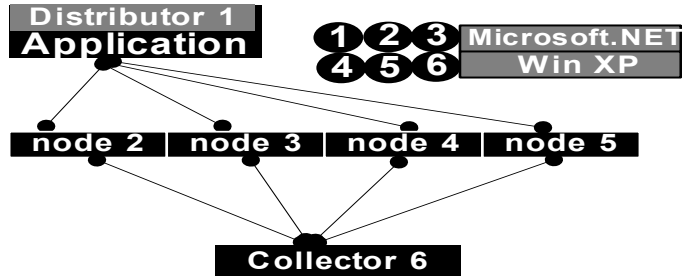

**Figure 9**   The peer-to-peer CryptEngine implementation architecture

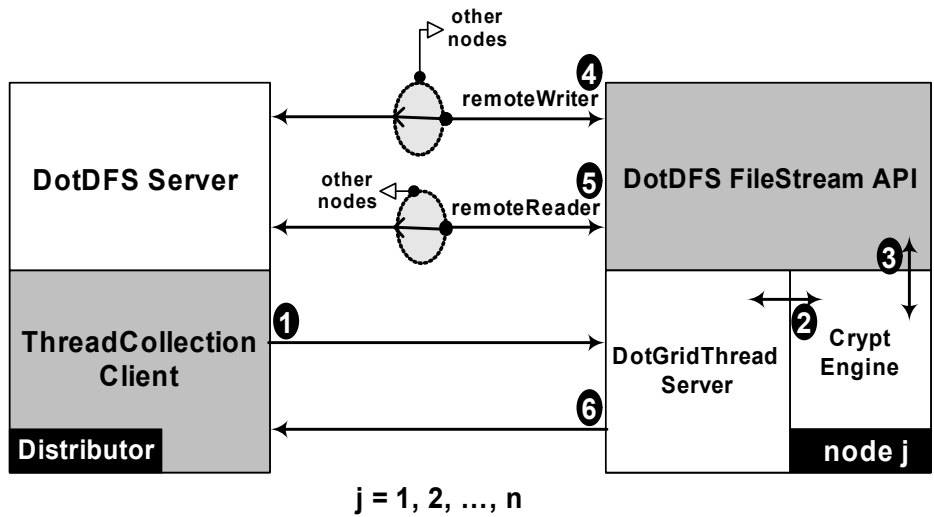

DotGrid.CryptEngine's most important features are as follows:

- Striping

  The architecture described above, lets every thread that encrypts fragments of file, store encrypted data in every node's storage system. Thus, such a parallelism, regarding the advances gained in 1 Gbps Ethernets that are connected to external networks at 10 Gbps or faster, can be used for storing encrypted files and striping them in a grid environment geographically.

- Memory usage

  In this remote multithreaded cryptography architecture, Distributor node distributes cryptographic tasks between other nodes, information are fetched via DotDFS in blocks and data in files are encrypted as file streams. This way, Distributor node and other nodes use the minimum memory and so very large files (*e.g.*, 1 GB files or greater) can be encrypted with high performance.

- High-throughput cryptography engine

  The DotGrid.DotDFS protocol along with its implementation has set up a high throughput distributed file system. Previously, for transmitting a 1GB file between two nodes we reached up to 623 Mbps speed (Abutalebi *et al.*, 2006), so CryptEngine can be used for large data encryption with high performance and throughput on a set of commodity computers in grid and cluster environments. Of course, this performance can be enhanced by using higher bandwidth networks and faster storage systems such as SCSI hard disks.

### 4.1.2 Experimental results

The testbed architecture for evaluation of CryptEngine performance is depicted in Figure 8. This testbed is a clustered grid environment containing six nodes. All machines have dual processors 1.5 GHz Pentium, 1 GB RAM and 80 GB IDE hard disk. DotDFS uses the NTFS file system for storing and fetching encrypted/decrypted data files in storage nodes on Windows XP OS. In two different tests, files of sizes 100, 200 and 400 MB have been encrypted by CryptEngine based on Rijndael and 3DES cryptographic algorithms. After encrypting any data file fragment with the size of BlockSize in a node, the node retransmits the encrypted data block to the Collector node by DotDFS as a file stream. This approach can be very suitable to encrypt and transfer petabytes of data files in Grid environments between two grids from a grid site to another, with the best throughput, efficiency and security. In these three tests, each file block length is 20 MB (BlockSize = 20 MB) and these symmetric cryptography algorithm's key and IV lengths are 128 bits. All tests have been performed in a LAN network with a bottleneck link of 100 Mbps. We repeated each test ten numbers of time with different number of nodes. Gained results are illustrated in Figures 10 and 11.

**Figure 10** DotGrid CryptEngine results for the Rijndael algorithm

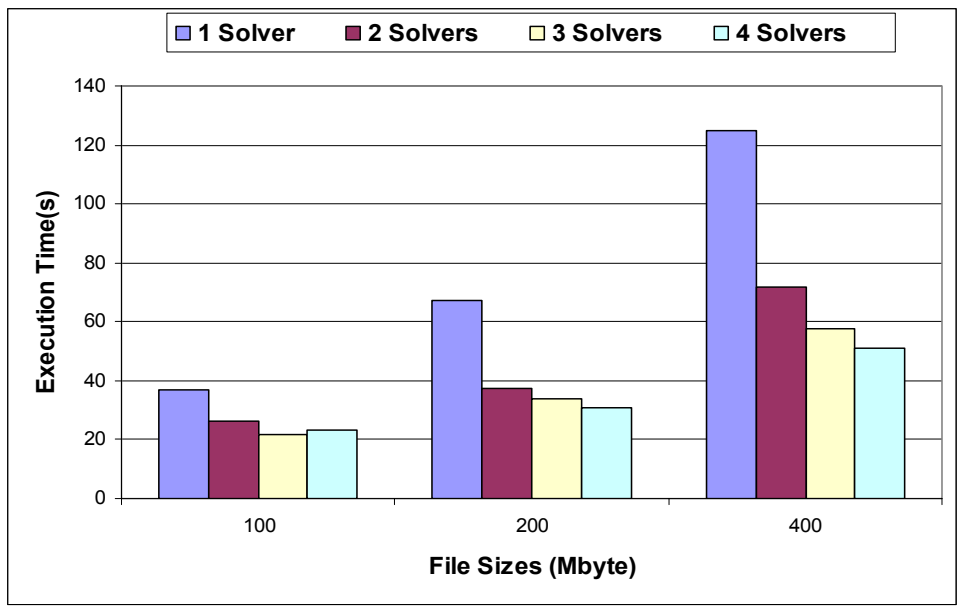

**Figure 11** DotGrid CryptEngine results for the 3DES algorithm

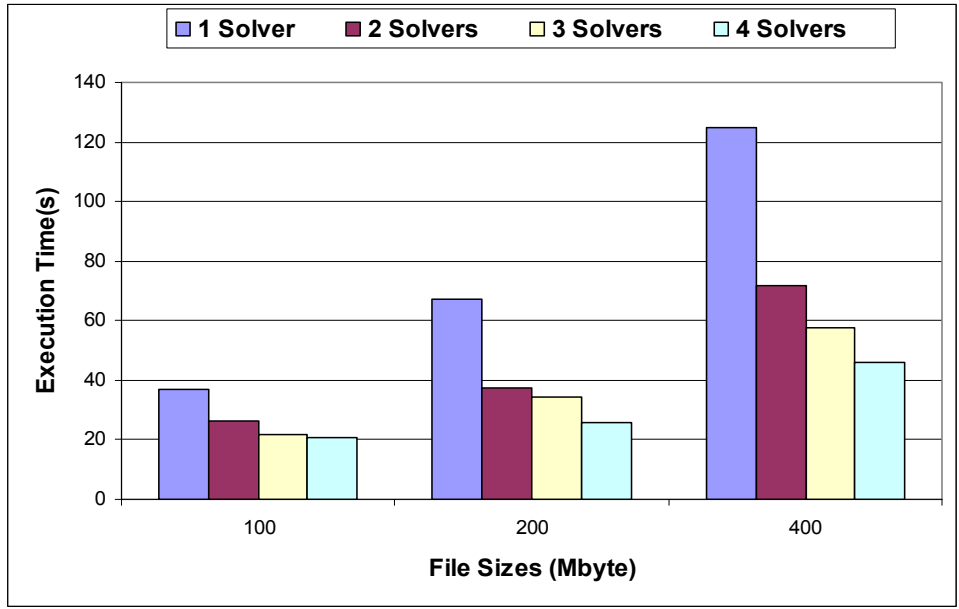

## 4.2 Comparison DotGrid with Alchemi

For comparing DotGrid to Alchemi, we implemented a typical application that is based on master-slaves architecture just like Alchemi (Luther *et al.*, 2005; Alchemi, 2007) middleware. The architectures of these two systems (DotGrid and Alchemi) are shown in Figure 12. The operating system for testing these two systems is Windows XP with Microsoft .NET Framework 1.1. All four machines are combined with dual processor 1.5 GHz Pentium, 1 GB RAM. In this typical application π number is calculated up to n digits (Bellard, 2006) utilising Single Programme, Multiple Data (SPMD) DotGrid model based on DotThreading service. In the Alchemi test, there is one manager and three executers. In the DotGrid test, the Distributor node is responsible for managing and distributing threads over the three remaining nodes. Both platforms' results are depicted in Figure 13.

**Figure 12** The DotGrid and Alchemi architecture testbed

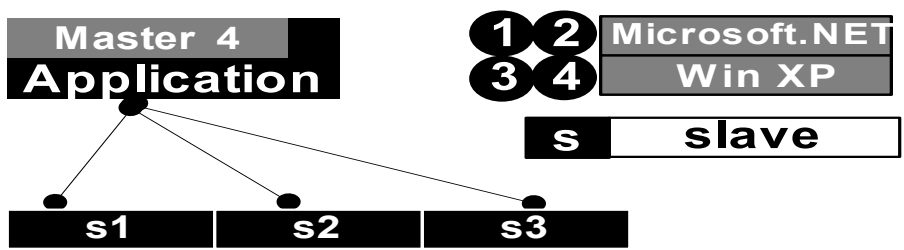

**Figure 13** Computation time for different number of digits of Pi in DotGrid and Alchemi

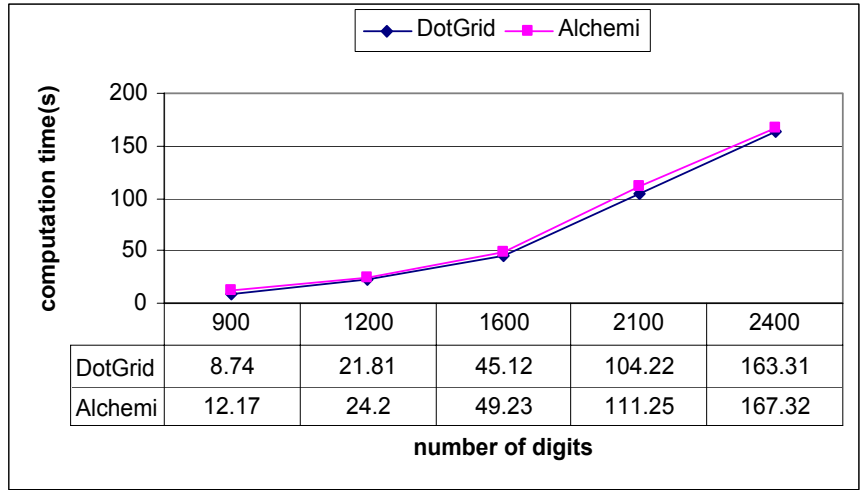

DotGrid results are better than those of Alchemi. Although implementing a grid like Alchemi is not the main mission of DotGrid, this example demonstrates the capability of DotGrid services and APIs for implementing a high-performance grid middleware. As it can be seen in Figure 13, results gained in calculation of number pi, depict that DotGrid platform performance is better than Alchemi middleware performance. This is mainly because of how these two systems have been implemented and the technologies used in kernel design of these two systems that effect conspicuously performance, efficiency, overhead and throughput of the systems. In this section, two main differences in architecture design of DotGrid and Alchemi that have caused performance improvement in DotGrid are considered:

- .NET Remoting

    Microsoft .NET Remoting (Obermeyer and Hawkins, 2001) creates an infrastructure for interacting objects in separate application domains. The .NET Remoting framework provides a set of services including remote object activation, lifetime support and communication channels that are responsible for message transmitting between two remote applications. Formatters are used for message encoding/decoding before they are transmitted via channels. The applications whose performance is critical can use binary encoding in TCP channels. Besides .NET Remoting provides XML encoding for interoperability with other remoting frameworks in HTTP channels based on the SOAP protocol. One of the most important features of .NET Remoting is hiding implementations of necessary infrastructures of message transmitting and complexities of calling remote methods. But these features that are used in .NET Remoting design, cause the same impacts on performance and data transmission between .NET Remoting clients and servers (just like Proxy Objects) (Obermeyer and Hawkins, 2001). Although Microsoft has done some efforts to correct these shortcomings (*e.g.*, JIT compiler and Execution Engine (EE) optimisations), Alchemi uses .NET Remoting for communication between its distributed grid components such as Executors, Grid Brokering Nodes, Managers

and Owners extensively. In contrast, DotGrid has developed all its services based on its open proposed protocols and low-level standard Berkeley sockets in order to overcome these overheads and promote its portability to other platforms such as Java and native code via standard C. All DotGrid protocols that state mechanisms for message passing and data transfer between grid nodes, are designed and optimised in such a way that the best performance is gained and optimal use of network bandwidth in TCP/IP sessions are reached. Thus, the most important reason for the DotGrid platform's high performance is not using .NET Remoting.

- Job data file transfers

    Alchemi lacks any kind of data grid capability. Alchemi encodes all binary and non-binary files (*e.g.*, .txt, .dll and .exe) based on Base64Encoding and then transfers them between nodes in order to transmit job files between executers, owners and managers. Encoding/decoding tasks that convert data from ASCII to binary and vice-versa, consumes huge CPU time especially for large data files and therefore kills computational time. On the other hand, overheads of this mechanism decrease system performance and speed. Besides, Alchemi cannot be used for the development of grid applications that have large – giga scale and higher – job files such as input, output and executing files because grid nodes memories are limited. DotGrid uses DotDFS as one of its data grid components. DotDFS not only has a high data throughput, but also makes using DotGrid in grid environments where petabytes of data are generated possible and even efficient. DotDFS proposes a streaming binary protocol for reaching high throughput in large data transmissions. All DotGrid job files between grid nodes are transmitted peer-to-peer in file streams while Alchemi transfers job files once from Owner to Manager and once again from Manager to Executors, that imposes another overhead on Alchemi performance especially for large files.

Finally, some other DotGrid and Alchemi differences are mentioned as follows:

- DotGrid provides a cross-platform infrastructure by its open protocols, its implementations and MySQL Database. Right now, DotGrid is executed on all platforms via MONO .NET, while it is impossible to port Alchemi to Linux because it uses Windows Services and Windows Forms along with MSSQL 2000 Database for setting up. For porting Alchemi, fundamental changes must be applied to its kernel. DotGrid architecture can be ported to native code with minor changes.

- Alchemi does not support TLS and GSI for secure data transmission between clients and managers. Besides, considering that Alchemi uses .NET Remoting, adding TLS or GSI to Alchemi kernel is very challenging (Toub, 2003). DotGrid uses DotSec for secure communications which not only support SSL3, TLS and GSI, but also express some open protocols so that all grid applications can use them based on their own policy. The DotSec protocol states a new Transport Security Layer architecture for reaching high throughput in data transfer in secure mode. Basically, DotSec has been designed and implemented based on DotDFS architecture.

*4.3 Linux DotDFS evaluation*

Our implementation of the DotDFS protocol discussed earlier supports all features in FTSM and DFSM modes. We performed our experiments in a LAN network with a bottleneck link of 512 Mbps and 0.02 ms Round Trip Time (RTT). Two machines used for clients and servers had Intel Celeron 2.8 GHz CPU, 512MB RAM and 80 GB IDE Hard Disk. Both machines had SUSE Linux 9.1. We used ReiserFS file system in Linux. MONO .NET Framework 1.1.8 was installed on Linux. We have developed a command-line DotDFS client utility by using our extensible DotDFS APIs. In all tests, we set the TCP buffer size and block size for disk IO to 256 KB. DotDFS uses DotSec for all transfer channels authentications. All achieved results are the means of 10 runs. We evaluated effectiveness of parallel TCP streams to increase total in SUSE Linux environment. Figure 14 depicts achieved results for a variety of parallel streams for both memory-to-memory (/dev/zero to /dev/null) tests and disk-to-disk tests scenarios. In disk-to-disk tests, we transferred a 1 GB file from client to server.

**Figure 14** Parallel throughput on LAN

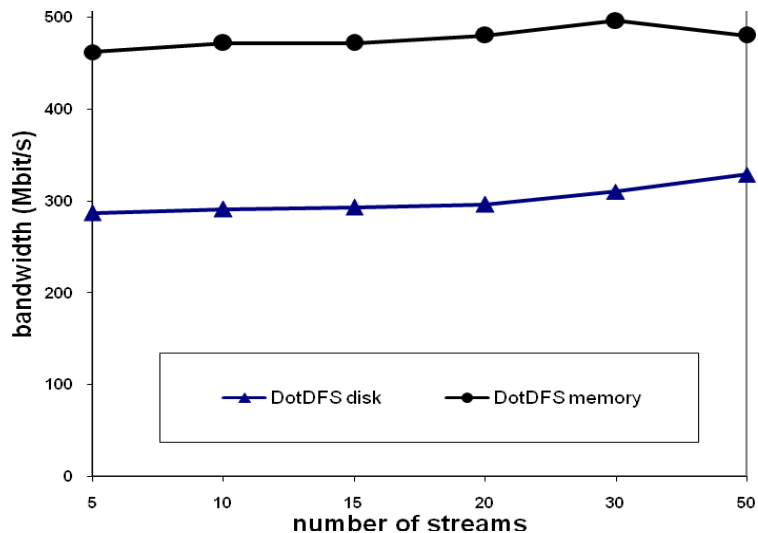

## 5 Conclusion and future work

We have developed DotGrid, an infrastructure for desktop grid environments that provides numerous services and APIs for developing next generations of grid computing middleware and distributed environments. CryptEngine architecture was explained in this paper as one of subapplications of DotGrid and is based on one of its models named hierarchical master-slaves topology. The CryptEngine architecture used for Symmetric and Asymmetric encryption demonstrates its power for numerous ranges of applications in grid and cluster computing. We are in the process of design and implementation of a Grid-based Cluster Parallel File System (GCPFS) for storing generated petabytes of data in grid environments. Our plan is to make it possible to use architectures and mechanisms mentioned in this paper for designing CryptEngine in pursuit of storing encrypted files in GCPFS infrastructure. Our other future working domains consist of extending current DotGrid services and APIs to build a smart and auto-controlled multiclustered middleware that will connect and manage a large number of heterogeneous DotGrid clusters distributed around the world for establishing a universal and unified grid environment in a hierarchical configuration. This approach will make considering and supporting some required complex technologies inevitable. These technologies are fault-tolerant, distributed load-balancing mechanisms, peer-to-peer system constraints and considerations such as NAT and Firewalls stated in Global Grid Forum (Bhatia, 2005) and system scalability. Also, we plan to interact and integrate DotGrid with Globus Toolkit (Foster, 2006) through .NET Platform Invoke. We intend to do this by creating our classified APIs. DotGrid will homogenise and unify the heterogeneous nature of distributed and grid computing environments.

## References

Abutalebi, A., Poshtkuhi, A., Ayough, L. and Hessabi, S. (2006) 'DotGrid: a .NET-based cross-platform grid computing infrastructure', *Proceedings of the IEEE International Conference on Computing and Informatics 2006 (ICOCI06)*, Malaysia, 6–8 June.

Alchemi (2007) 'Alchemi .NET home page', http://www.alchemi.net/.

Allcock, B., Mandrichenko, I. and Perelmutov, T. (2005a) 'GridFTP v2 protocol description', http://www.ggf.org/documents/GFD.47.pdf.

Allcock, W. (2003) 'GridFTP: protocol extensions to ftp for the grid', http://www.ggf.org/documents/GFD.20.pdf.

Allcock, W., Bresnahan, J., Kettimuthu, R., Link, M., Dumitrescu, C., Raicu, I. and Foster, I. (2005b) 'The globus striped GridFTP framework and server', *Proceedings of Super Computing 2005 (SC05)*, November.

Anderson, D., Cobb, J., Korpela, E., Lebofsky, M. and Werthimer, D. (2002) 'SETI@home: an experiment in public-resource computing', *Communications of the ACM*, USA: ACM Press, November, Vol. 45, No. 11.

Baghdadi, Y. and Al-Rawahi, N. (2007) 'An architecture and a method for web services design: towards the realization of service-oriented computing', *International Journal of Web and Grid Services 2006*, Vol. 2, No. 2, pp.119–147.

Belalem, G. and Slimani, Y. (2007) 'A hybrid approach to replica management in data grids', *International Journal of Web and Grid Services*, Vol. 3, No. 1, pp.2–18.

Bellard, F. (2006) 'Computation of the n'th digit of pi in any base in O(n^2)', http://fabrice.bellard.free.fr/pi/pi_n2/pi_n2.html.

Bhatia, K. (2005) 'Peer-to-peer requirements on the open grid services architecture framework', *OGSAP2P Research Group in Global Grid Forum (GGF)*, GFD-I.049, 12 July, http://www.ggf.org/documents/GFD.49.pdf.

Burke, J., McDonald, J. and Austin, T. (2000) 'Architectural support for fast symmetric-key cryptography', *Advanced Computer Architecture Laboratory University of Michigan.*

Buyya, R. and Venugopal, S. (2004) 'The Gridbus toolkit for service-oriented grid and utility computing: an overview and status report', *Proceedings of the First IEEE International Workshop on Grid Economics and Business Models*, USA.

Chien, A., Calder, B., Elbert, S. and Bhatia, K. (2003) 'Entropia: architecture and performance of an enterprise desktop grid system', *Journal of Parallel and Distributed Computing*, USA: Academic Press, May, Vol. 63, No. 5.

Clark, J. (2003) 'Calling Win32 DLLs in C# with P/Invoke', *MSDN Magazine*, The Microsoft Journal for Developers, July, http://msdn.microsoft.com/msdnmag/issues/03/07/NET/default.aspx.

CryptographyWorld.com (2006) 'The cryptography guide: triple DES', http://www.cryptographyworld.com/des.htm.

Daemen, J. and Rijmen, V. (2002) 'The design of Rijndael, AES', *The Advanced Encryption Standard*, Springer-Verlag.

Dierks, T. and Allen, C. (1999) 'The TLS protocol version 1.0', *IETF*.

Dongara, P. and Vijaykumar, T.N. (2003) 'Accelerating private-key cryptography via multithreading on symmetric multiprocessors', *IEEE International Symposium on Performance Analysis of Systems and Software (ISPASS)*, March.

ECMA-334 (2007) 'ECMA-334: C# language specification', http://www.ecma-international.org/publications/standards/Ecma-334.htm.

ECMA-335 (2007) 'Common Language Infrastructure (CLI)', http://www.ecma-international.org/publications/techreports/E-TR 084.htm.

El-Rewini, H. and Abd-El-Barr, M. (2005) 'Parallel programming in the parallel virtual machine', *Advanced Computer Architecture and Parallel Processing*, ISBN: 0-471-46740-5, John Wiley.

Foster, I. (2006) 'Globus toolkit version 4: software for service-oriented systems', *IFIP International Conference on Network and Parallel Computing*, Springer-Verlag LNCS 3779, pp.2–13.

Foster, I., Kesselman, C., Nick, J. and Tuecke, S. (2002) 'The physiology of the grid: an open grid services architecture for distributed systems integration', *Open Grid Service Infrastructure WG*, Global Grid Forum, 22 June.

Foster, I., Kesselman, C. and Tuecke, S. (2001) 'The anatomy of the grid enabling scalable virtual organizations', *International J. Supercomputer Applications*, Vol. 15, No. 3.

GCG (2007) 'Grid Computing Group at Uva', http://www.cs.virginia.edu/~humphrey/GCG.htm.

Grimshaw, A.S. and Wulf, W.A. (1997) 'The legion vision of a worldwide virtual computer', *Communications of the ACM*, January, Vol. 40, No. 1, pp.39–45.

Huber, V. (2001) 'UNICORE: a grid computing environment for distributed and parallel computing', *Proceedings PaCT*, Springer Verlag, Vol. 2127, pp.258–266.

Interoperating (2004) 'Interoperating with native libraries through MONO .NET framework', http://www.mono-project.com/Interop_with_Native_Libraries.

Jonsson, J. and Kaliski, B. (2003) 'RSA cryptography specifications version 2.1', *IETF*, RFC 2437.

Keahey, K., Ripeanu, M. and Doering, K. (2003) 'Dynamic creation and management of runtime environments in the grid', *Workshop on Designing and Building Grid Services*, GGF-9.

Linn, J. (1997) 'Generic security service application program interface, version 2', *Internet RFC 2078*.

Luther, A., Buyya, R., Ranjan, R. and Venugopal, S. (2005) 'Alchemi: a .NET-based enterprise grid computing system', *Proceedings of the 6th International Conference on Internet Computing (ICOMP'05)*, Las Vegas, USA, 27–30 June.

Microsoft Corporation (2007) '.NET framework home', http://msdn.microsoft.com/netframework/.

MONO .NET (2007) 'MONO .NET project home page', http://www.mono-project.com/.

Obermeyer, P. and Hawkins, J. (2001) *Microsoft .NET Remoting: A Technical Overview*, Microsoft Corporation, July.

Platform Computing (2007) http://www.platform.com.

Poshtkuhi, A., Abutalebi, A., Ayough, L. and Hessabi, S. (2006) 'DotGrid: a .NET-based infrastructure for global grid computing', *Proceedings of the CCGrid06 IEEE International Workshop on Grid Testbeds*, Singapore, 18–19 May.

Rivest, R. (1998) 'A description of the RC2(r) encryption algorithm', *Internet RFC 2268*.

Setiawan, A., Adiutama, D., Liman, J., Luther, A. and Buyya, R. (2004) 'GridCrypt: high performance symmetric key using enterprise grids', *Proceedings of the 5th International Conference on Parallel and Distributed Computing, Applications and Technologies*.

Sullivan, W.T., Werthimer, D., Bowyer, S., Cobb, J., Gedye, D. and Anderson, D. (1997) 'A new major SETI project based on project Serendip data and 100,000 personal computers', *Proceedings of the 5th International Conference on Bioastronomy*.

Thain, D., Tannenbaum, T. and Livny, M. (2003) 'Condor and the grid', in F. Berman, A.J.G. Hey and G. Fox (Eds.) *Grid Computing: Making The Global Infrastructure a Reality*, John Wiley.

Toub, S. (2003) 'Secure your .NET remoting traffic by writing an asymmetric encryption channel sink', *MSDN Magazine*, The Microsoft Journal for Developers, July, http://msdn.microsoft.com/msdnmag/issues/03/06/NETRemoting/toc.asp.

United Devices (2007) 'Grid MP™: the award-winning technology for enterprise application virtualization', http://www.ud.com/products/gridmp.php.